\input{aipcheck}

\documentclass[
    ,final            
  ]
  {aipproc}

\layoutstyle{6x9}

\def\sigtot{\sigma_{total}}

\begin{document}

\title{QCD and total cross-sections : photons and hadrons}


\author{R.M. Godbole}{
  address={Centre for High Energy Physics, Indian Institute of Science,
 Bangalore, 560012, India}
}

\author{A. Grau}{
  address={Departamento de F\'\i sica Te\'orica y del Cosmos 
Universidad de Granada, 18071 Granada, Spain}
}

\author{G. Pancheri}{
  address={INFN Frascati National Laboratories, Via Enrico Fermi 40, I-00044 Frascati, Italy}
}
\author{Y.N. Srivastava}{
 address={INFN and Physics Department, University of Perugia,
Via A. Pascoli, I-06123 Perugia, Italy}
  }

\begin{abstract}
In this contribution, we discuss  a total cross-section model which can be applied to both photon and purely hadronic processes. We find that the model can reproduce photo-production cross-sections, as well as  extrapolation of $\gamma^* p$ processes to $\gamma p$ using Vector Meson Dominance models, with minimal modifications from the proton case. 



\end{abstract}

\maketitle


\section{Introduction}

 All total cross-sections are seen to rise asymptotically \cite{PDG}. For all processes,  the rise  starts in the region $\sqrt{s}\approx 10\div 50\  GeV$. It has long been advocated that the rise is due to the onset of perturbative parton-parton collisions \cite{halzencline} and a large amount of modelling has been developed to describe the energy dependence of the total cross-section. 
A compilation of total cross-section data  \cite{gampnew} from purely hadronic processes \cite{PDG,dataproton} together with the available ones from photon processes \cite{h1,zeus,datagg} shows that  while photoproduction data could  accomodate a rise similar to the pure hadronic  processes,   this may not be so for purely photon processes. Indeed we have shown this not to be true for the case of $\gamma \gamma$ scattering \cite{withalbert}.

If one ignores the limits imposed by the Froissart-Martin  bound, namely $\sigtot \le  C\log^2{s}$, a phenomenologically successful model for  the energy behaviour of  total cross-sections for different processes is given by  the   expression proposed in \cite{DL}, namely
\begin{equation}
\sigma_{tot}^{AB}=X_{AB}s^{-\eta}+Y_{AB}s^{\epsilon}
\end{equation}

with $\epsilon \approx 0.08\div 0.09$.  The above equation however implies that all the cross-sections 
 should have the same power law behaviour.

In order to develop a model which might  go beyond the low energy description, we notice that the range of energy  of total cross-section measurements can be divided into four different regions:
\begin{itemize}
\item the resonance region which depends on the quantum numbers of the s-channel process
\item the Regge region, where analiticity, crossing and unitarity predict an energy behaviour  with a decreasing power law, i.e. $\sigtot \simeq s^{-\eta}$, and $\eta\approx 0.5$ and whose details  depend on the quantum numbers of the t-channel process
\item a region with  the onset of the rise and where $\sigtot$ starts rapidly to grow,  possibly with a power law type behaviour
\item the truly asymptotic region where the cross-section should obey the Martin-Froissart bound.
\end{itemize}
The first two regions are related in their energy behaviour through Finite Energy Sum Rules (FESR) \cite{fesr}, while the last two  are dominated by the dynamics of perturbative QCD processes. In the model we have developed \cite{ourlast} and which shall be described in this contribution,  we focus on the high energy (rising) part, and describe the rise by perturbative gluon-gluon scattering, with saturation effects, which  lead to a satisfaction of the Froissart bound  due to soft gluon emission. According to our model, infact, resummation of gluons in the infrared region is the major origin of the transformation of the rise from a power law to a $\log{s}$ or $\log^2{s}$ type behaviour.

\section{The Bloch-Nordsieck model for the total cross-section}
We work in the eikonal representation, for which
 \begin{equation}
\sigma^{\gamma p}_{tot} =
2 P_{had}^{AB}\int d^2 {\vec b}[1-e^{-n(b,s)/2}] 
\label{sigtot}
\end{equation}
with $P_{had}^{AB}=1$ for purely proton processes, otherwise related to the probability that a photon behaves like a hadron, for which a matter distribution is expected. We use  $P_{had}^
{\gamma \gamma}=[P_{had}^{\gamma }]^2$ and put $P_{had}^{\gamma }\equiv P_{had}=1/240$ from Vector Meson Dominance.

In Eq. \ref{sigtot} the real part of the eikonal function has been approximated to zero, and the imaginary part is given by the average number of inelastic collisions $n(b,s)$. To calculate this quantity, we distinguish between collisions with an  outgoing parton  with $p_t \ge p_{tmin}$, with  $p_{tmin}$ in a region where perturbative QCD calculations can be used, and all other collisions, i.e. we write
\begin{equation}
n(b,s)=n_{soft}(b,s) + n_{hard}(b,s)
\end{equation}
and calculate $n_{hard}(b,s)$ from QCD, using QCD mini-jets to drive the rise. For photoproduction  processes we have
\begin{eqnarray}
\label{nbs}
n^{\gamma p}(b,s)=n_{soft}^{\gamma p}(b,s) +n^{\gamma p}_{hard}(b,s) = n_{soft}^{\gamma p}(b,s) + A(b,s) \sigma_{jet}^{\gamma p}(s)/P_{had}
\end{eqnarray} 
with  $n_{hard}$ including all outgoing parton processes with $p_t>p_{tmin}$. The parameter $p_{tmin}$ is a cut-off imposed on the jet cross-section, namely 
\begin{equation}
\sigma^{AB}_{\rm jet} (s,p_{tmin})= 
\int_{p_{tmin}}^{\sqrt{s}/2} \!\! d p_t \int_{4
p_t^2/s}^1 \! \!\! d x_1  \int_{4 p_t^2/(x_1 s)}^1 \! \! \! \! d x_2 
 \sum_{i,j,k,l}
f_{i|A}(x_1,p_t^2) f_{j|B}(x_2,p_t^2)
  \frac { d \hat{\sigma}_{ij}^{ kl}(\hat{s})} {d p_t} \ \ \ \ \ \
  \end{equation}
where $A$ and $B$ are the colliding hadrons or photons, in this case $A-proton, B-\gamma$. By construction,  this cross-section  depends on the particular parametrization of the DGLAP 
evoluted parton distribution functions (PDFs). Phenomenology of the early rise in proton-proton suggests $p_{tmin} \approx 1\  GeV$, and these low energy jets are called {\it mini-jets}.
Because the jet cross-sections are calculated using actual  photon densities, 
which themselves give the probability of finding a given quark or gluon in a 
photon, $P_{had}$ needs to be  canceled out in $n_{hard}$. 

 In Eq.\ref{nbs}  the impact parameter dependence has been factored out:  this is  an approximation which could be relaxed in the future.
  There exist many models in the literature to describe the impact parameter distribution of matter in the colliding particles, the earliest of such models based on convolution of the  particles form factors. The problem with these  models is the difficulty to extend them to  the photon, although Vector Meson Dominance (VMD) might suggest to use a meson-like description, namely a monopole functional dependence, with an {\it ad hoc} scale. There exist also more fundamental recent studies from perturbative Reggeon calculus \cite{bartels}.
\subsubsection{The impact parameter distribution}
In our model the matter distribution in  two colliding hadrons or hadron-like particles (for the case of the photon in its interactions with matter) are determined by the $k_t$ distribution of soft gluons emitted in the interaction \cite{ourPRD}. As the partons move through the field of the other colliding quarks and gluons, they are deviated from their collinear path through soft gluon emission.  The Fourier transform of the resultant resummed $k_t$-distribution gives the energy-dependent impact parameter function to enter the eikonal. We use
\begin{equation}
 A^{AB}_{BN}(b,s) =
{\cal N} \int d^2 {\bf K}_{\perp} {{d^2P({\bf K}_\perp)}\over{d^2 {\bf K}_\perp}} 
 e^{-i{\bf K}_\perp\cdot {\bf b}} \nonumber 
 = {{e^{-h( b,q_{max})}}\over
 {\int d^2{\bf b} e^{-h(b,q_{max})} }}\equiv A^{AB}_{BN}(b,q_{max}(s)).\ \ \
 \label{Eq:abn}
 \end{equation}
 The function $A^{AB}_{BN}$ is normalized to 1 and is obtained from
the Fourier transform  of the soft gluon resummed transverse 
momentum distribution. This impact parameter distribution is energy dependent through the function
\begin{equation}
h( b,q_{max}(s))  = 
\frac{16}{3}\int_0^{q_{max}(s) }
{{dk_t}\over{k_t}} 
 {{ \alpha_s(k_t^2) }\over{\pi}} \left(\log{{2q_{\max}(s)}\over{k_t}}\right)\left[1-J_0(k_tb)\right]
\label{hdb}
\end{equation}
which is defined by $q_{max}(s)$, the maximum transverse momentum allowed to single gluon emission, averaged over the parton-parton cross-sections, as described in \cite{gampnew}.  To fully include in the model the very low momentum gluons, emitted in indefinite number during the interaction, we have made an ans\"atz as to the behaviour of the strong coupling constant $\alpha_s(k_t^2)$ 
with
\begin{equation}
 \alpha_s(k_t)={{12 \pi}\over{33-2N_f}} {{p}
\over{\ln[1+p({{k_t}\over{\Lambda}})^{2p}]}}
\approx 
constant \times \left({{\Lambda}\over{k_t}}\right)^{2p}\ \ \ \ k_t\to 0
\label{alphas}
\end{equation}
with $p<1$ for the soft gluon integral to converge.
\section{Photoproduction total cross-section}
\begin{figure}
  \includegraphics[height=0.7\textheight]{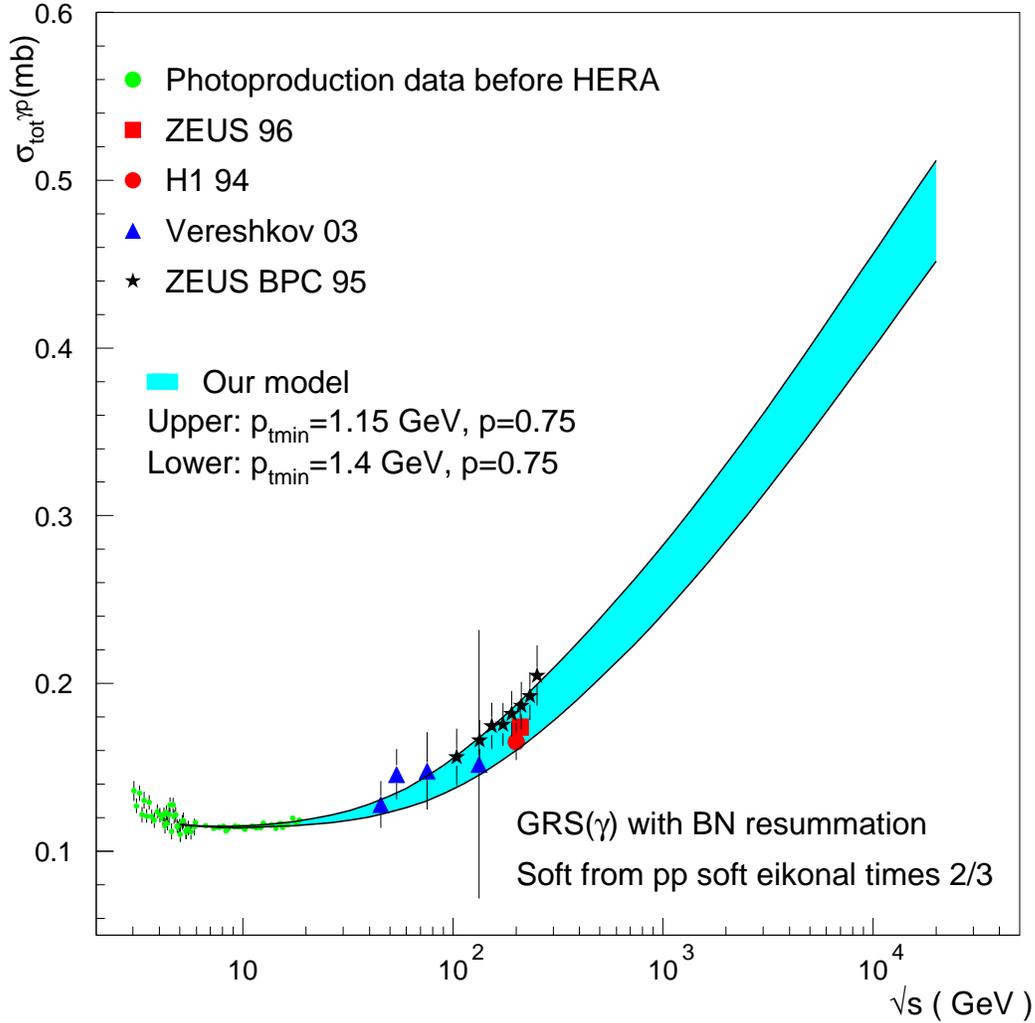}
  \caption{Description of $\gamma p$ data from the  Bloch Nordsieck model described in the text }
  \label{gampgrs115}
\end{figure}

We show in Fig. \ref{gampgrs115} the results of this model when applied to photoproduction data, and to a set of $\gamma^* p$ data from ZEUS BPC, extrapolated to $Q^2=0$ using Generalized Vector Meson Dominance \cite{haidt,bernd,bpc}. In the figure, the upper end of the band corresponds to the model with the same parameters found to give a good fit to proton-proton and proton-antiproton scattering, namely GRV densities for the proton, $p_{tmin}=1.15\  GeV$, $p=0.75$. The lower edge corresponds to a higher value of $p_{tmin}$, as indicated. Curves for other values of the parameters and different densities for quarks and gluons are described in ref. \cite{gampnew}.


In conclusion, we have applied  to photoproduction our  QCD model, originally developed to describe total cross-sections for purely hadronic processes, obtaining  a good description of existing data, with minimal changes in the model parameters.


\begin{theacknowledgments}
R.G. aknowledges support from the Department of Science and Technology, India, under the J.C. Bose fellowship. G. P.  thanks   the  Laboratory for Nuclear Science of the Massachussetts Institute of Technology Laboratory   for hospitality 
while this work was being written.
This work has been partially supported by MEC (FPA2006-\-05294) and  Junta
de 
Andaluc\'\i a (FQM 101 and FQM 437).
\end{theacknowledgments}

\end{document}